\documentclass{mem}
\usepackage{natbib}\usepackage{txfonts}\usepackage{balance}
\usepackage{graphicx}
\usepackage[a4paper]{hyperref}
\idline{75}{282}
\begin{document}
\def\teff{$T\rm_{eff }$}
\def\kms{$\mathrm {km s}^{-1}$}

\title{
Anomalous globular clusters: 
insights from neutron capture elements abundances.}

   \subtitle{}

\author{
A. F. \, Marino\inst{1,2} 
         }

  \offprints{A. F. Marino}

\institute{
Max Planck Institute for Astrophysics, Postfach 1317,
D-85741 Garching, Germany
\email{amarino@MPA-Garching.MPG.DE}
\and
Research School of Astronomy \& Astrophysics, Australian National University, Mt Stromlo Observatory, via Cotter Rd, Weston, ACT 2611, Australia
\email{amarino@.mso.anu.edu.au}
}

\authorrunning{Marino }
\titlerunning{Neutron-capture elements in anomalous GCs}

\abstract{
Thanks to the large amount of spectroscopic and photometric
data assembled in the last couple of decades, the assumption
that all globular clusters (GCs) contain a simple mono-metallic
stellar population has been modified. 
Besides the common variations in the elements created/destroyed in the
H-burning processes, spreads and/or multi-modalities in heavier
elements have been detected in a few objects.
Among the most remarkable chemical inhomogeneity in these {\it anomalous} objects 
is the internal variation in the neutron-capture ($n$-capture)
elements, that can provide some information about the
material from which stars were born.
I report a summary of the chemical pattern observed in GCs where variations in $n$-capture
have been detected, and the connection between these chemical features and
the distribution of stars along the color-magnitude diagrams in the
context of the lively debate on multiple stellar populations.

}
\maketitle{}

\section{Introduction}

In recent years, observational evidence, both
from high resolution spectroscopy and from photometry, has established
that GCs can host more than one stellar
population.
Nearly all the GCs studied with a good statistics show internal variations in
the elements involved in the H-burning reactions, e.g. C, N, O, Na, Mg,
and Al (\cite{car09}).
In these clusters, that we may call {\it normal} GCs, stellar abundances
of elements heavier than those affected by H-burning resemble 
the halo field compositions at similar metallicities, and show 
internal consistency within observational errors.

Recent spectroscopic studies have revealed some chemical anomalies,
i.e. some GCs have variations not only in light elements, but also in the bulk
heavy element content, and significant metallicity
dispersions.
In some cases, these chemical anomalies are connected to peculiar
distribution of stars along  color-magnitude diagram (CMD),
e. g. split/broad sub-giant branches (SGBs)

GCs proven to be {\it anomalous} include NGC 6656 (M22, \cite{mar09}, \cite{dac09}; NGC~2419, \cite{coh10}; Terzan~5, \cite{fer09};
NGC~1851, \cite{yon08}; \cite{car11}), and
the Sagittarius dwarf galaxy central cluster M54 (\cite{sl95}; \cite{car10}). 
All these objects share superficial similarities with the most massive
and peculiar
GC $\omega$~Cen (e.g. \cite{dcm11}), whose huge metallicity variations have been
known since the 1970s (e.g. \cite{dw67}, \cite{fr75}, \cite{ndc95},
\cite{sk96}; and more recently \cite{jp10}; \cite{mar11b}).
In addition, some of these {\it anomalous} clusters show intriguing chemical star-to-star
variations in the abundance of the $n$-capture elements relative to Fe.

In the following I discuss the observational scenario for {\it
  anomalous} GCs, focusing on the two cases of M22 and $\omega$~Cen, with
connections between the chemical abundances of the stars and their
position along the  CMD.
Preliminar results for the case of 47~Tuc, where a
split SGB has been recently detected, have been also discussed in the
context of the chemistry-multiple SGBs connections.

\section{$n$-capture element variations in GCs}

The study of the star-to-star $n$-capture elements variations
provides fundamental knowledge about the history and nature of
nucleosynthesis in {\it anomalous} GCs. 
Indeed, $n$-capture elements can be synthesised
from the rapid ($r$) and slow ($s$) $n$-capture processes, expected to
occur in different stellar environments, i.e. in
explosive environments the former (e.g. \cite{was00}), and in long-lived low
and intermediate-mass stars the latter (e.g. \cite{bgw99}, \cite{kar12}).
In particular, abundance variation in elements predicted to be almost entirely
produced in $r$ (e.g. Eu 97\% $r$, \cite{sim04}) or $s$ processes can give potential
insights about the chemical enrichment in {\it anomalous} GCs.

In the multi-variegate zoo of {\it anomalous} GCs, there are objects
representative of both $r$ and $s$-capture element enrichment.
\cite{sne97} have shown that the GC M15 shows a Ba-Eu
correlation, with evidence for a bi-modal distribution on the [Ba/Eu]
vs. [Eu/Fe] plane (see their Fig.~6). 
Although in principle, $r$ processes can contribute to the Ba production,
the correlation with Eu points towards a main contribution from $r$-processes.

Omega~Cen, NGC~1851, and M22 show different degrees of variations in $n$-capture
elements, but, at odds with M15, the [Eu/Fe] abundance does not vary
within errors. 
In fact, successive stellar generations in these objects may
have formed from material processed in low-mass stars asymptotic-giant
branch (AGB) stars. 
A summary of the chemical pattern of $\omega$~Cen and M22 is the
subject of next sections.

\subsection{M22}\label{m22S}
For a long time $\omega$~Cen was considered the unique GC with
internal overall metallicity variations.
Thus the recent discovery (\cite{mar09}) of an
intrinsic Fe variation in M22 confirmed from high-resolution
spectroscopy was surprising. 
The most striking chemical feature of M22 is the bimodality in
the content of elements mainly produced in the $s$ elements
(\cite{mar09}, \cite{mar11a}, \cite{rms11}). 

\begin{figure}[t!]
\resizebox{\hsize}{!}{\includegraphics[clip=true]{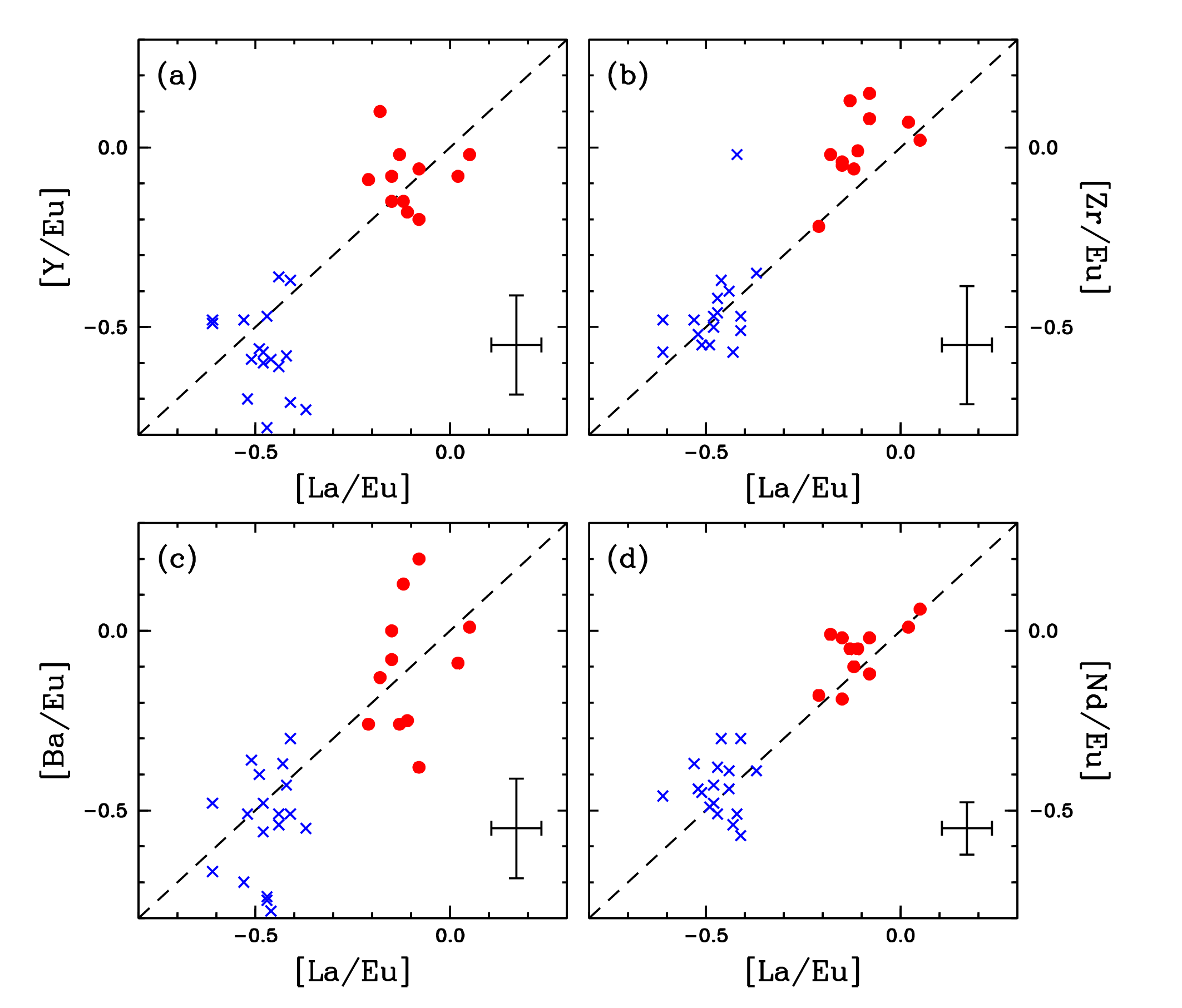}}
\caption{\footnotesize
Y, Zr, Ba, and La abundances relative to Eu as a function of [La/Eu]
in M22 from \cite{mar11a}. In each panel, the dashed line represents equality of the
displayed abundance ratios.  
Filled circles are used for stars with $s$-process enhancements and
$\times$ symbols are for stars without such enhancements.
}
\label{m22}
\end{figure}

\begin{figure}[t!]
\resizebox{\hsize}{!}{\includegraphics[clip=true]{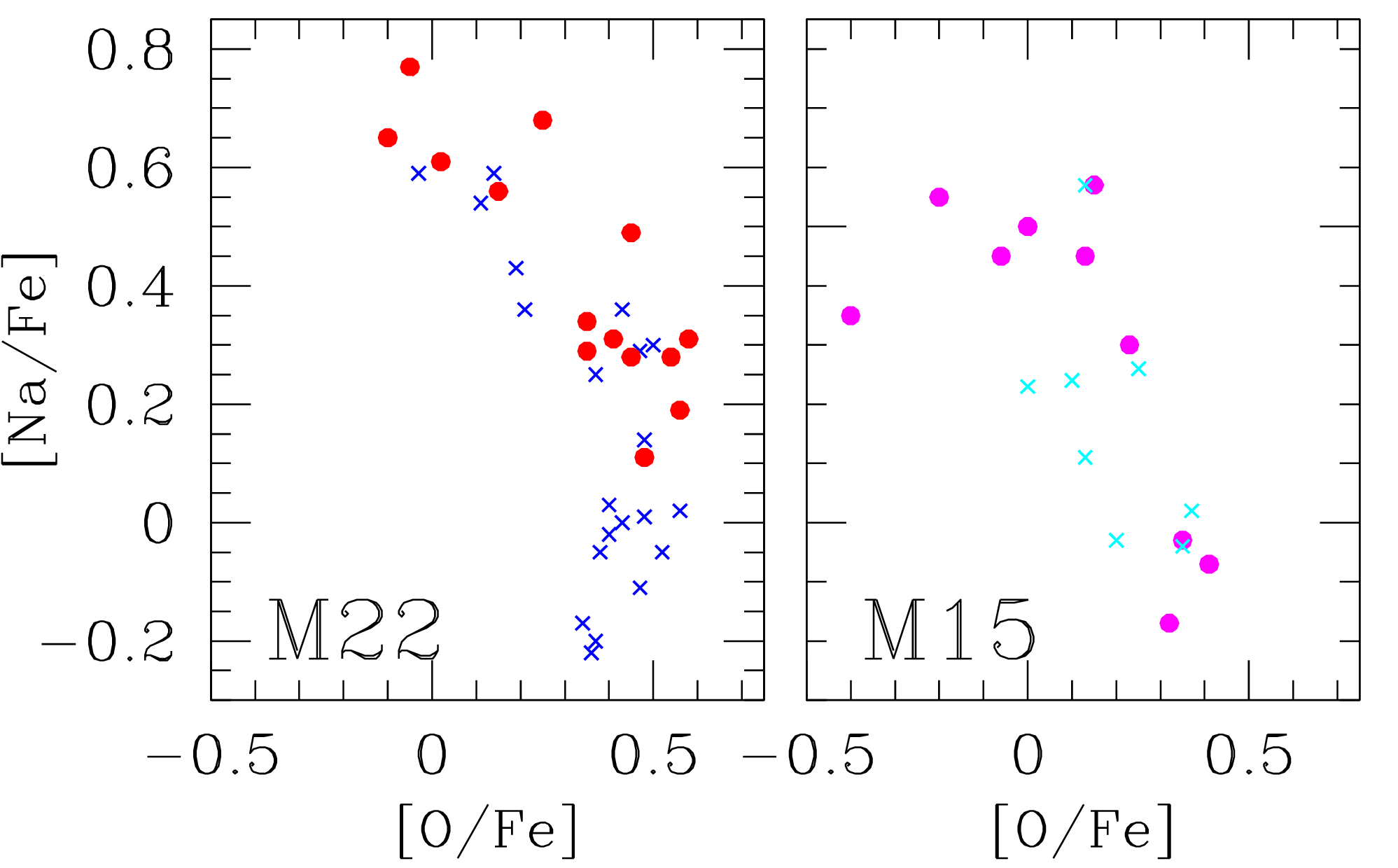}}
\caption{\footnotesize 
Na-O anticorrelation for M22 $s$-rich and $s$-poor stars from \cite{mar11a}
(symbols are as in Fig.~\ref{m22}), and for stars in
M15 (from \cite{sne97}). For the M15 Sneden et al. sample, cyan points
represent stars with [Eu/Fe]$<$0.50 dex, and magenta points stars with [Eu/Fe]$>$0.50 dex.
}
\label{m22m15}
\end{figure}

As shown in Fig.~\ref{m22} (from \cite{mar11a}) there is a bimodality in the distribution
of $s$-process elements in M22, while the $r$-process Eu is constant
within our observational errors.
A stellar group is enriched in $s$-process abundances ($s$-rich) with respect to
the other $s$-poor stars  (see \cite{mar11a} for a definition).
This bimodality corresponds to a different C+N+O content
(\cite{mar11a}, \cite{aab12}). 
Interestingly, each $s$-group individually defines a Na-O
anticorrelation (left panel in Fig.~\ref{m22m15}), suggesting that both have suffered from the same
enrichment occurred in normal GCs.
A spread in O and Na is also present in each stellar
group in M15, that, at odds with M22, shows evidence for an 
enrichment due to $r$-processed material (\cite{sne97}; right panel of
Fig.~\ref{m22m15}).

\subsection{Omega Centauri}

The complex multiple stellar population phenomenon in $\omega$~Cen
manifests in an intricate chemical pattern. 
The understanding of the chemical enrichment history of this object is
challenging and requires the knowledge of the chemical composition of
its hosted (maybe discrete) stellar populations.

The large spread in Fe in this cluster is known since a long time. 
However, in recent years, thanks to the spectroscopic homogenous
analysis of large samples of stars in this GC, at mid and high
resolution, it has been possible to study in more details the chemical
patterns of stars at different metallicities (e.g. \cite{jp10}, \cite{mar11b}).

\begin{figure}[t!]
\resizebox{\hsize}{!}{\includegraphics[clip=true]{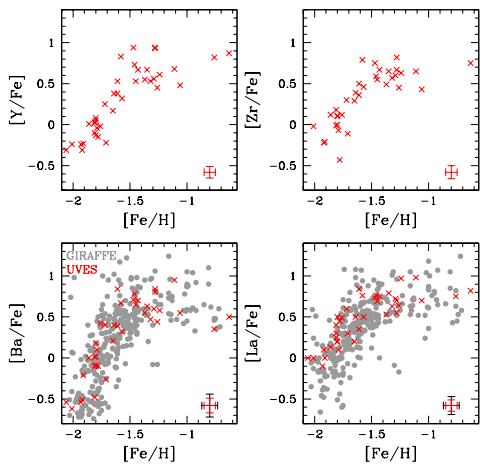}}
\caption{\footnotesize
Y, Zr, Ba, and La abundances relative to Fe as a function of
[Fe/H] in $\omega$~Cen. Gray points are GIRAFFE data (from
\cite{mar11a}), red points are UVES data (from \cite{mar13a}, in prep.). 
}
\label{omega}
\end{figure}

The run of the $n$-capture elements Ba and La as a function of metallicity from
the \cite{mar11b} GIRAFFE  dataset is shown
in Fig.~\ref{omega} (lower panels). 
Super-imposed to this dataset is a sample of stars studied at higher
resolution with UVES. For this sample the abundances of Y
and Zr have also been measured (\cite{mar13a}).
All these elements show a range larger than 1 dex, and 
a clear growth with metallicity up to [Fe/H]$\sim -1.5$; for higher
metallicities the distribution is flat (see also \cite{ndc95}, \cite{sm00},
\cite{jp10}).
 
Similarly to M22, variations in 
C+N+O abundances have been detected (\cite{mar12a}) in $\omega$~Cen.
The a Na-O anticorrelation is present along almost  the entire metallicity
range with the exception of the most metal poor and
most metal rich stars (\cite{mar11b}, \cite{jp10}).

To explain the intricate chemical pattern observed in $\omega$~Cen,
\cite{dan11} suggested that 
successive generations of increasing metallicity could form from
massive AGB ejecta diluted with the in-falling iron-enriched pristine
matter. This is predicted to occur until the diluting material is exhausted. Later
on, the most metal rich stars in the cluster may have been formed directly from
the massive AGB ejecta, and shows a direct Na-O correlation (as
suggested by the data sample presented in \cite{mar11b} and \cite{jp10}).
However, as \cite{dan11} pointed out the large growth in the abundances of
$s$-process elements observed in this peculiar cluster remains difficult to be understood
within a chemical evolution model.

\section{Connections with the CMD}

The {\it anomalous} GCs have been extensively photometrically
investigated.
Photometry from {\it Hubble Space Telescope} has revealed that M22 and
NGC~1851, that are chemically similar (\cite{mar09}, \cite{yon08},
\cite{car11}, \cite{lar12}), show a  split SGB (\cite{mil08}, \cite{pio12}). 

The analysis of GIRAFFE spectra of M22 SGB stars presented in
\cite{mar12b} has demonstrated that the two SGBs are 
populated by the two $s$ stellar groups (see Sect.~\ref{m22S}), 
and the split can be ascribed to the different CNO
content observed on the RGB (\cite{mar11a}), as predicted by
\cite{cas08} and \cite{ven09}. 

\begin{center}
\begin{figure*}[t!]
\includegraphics[width=4cm,height=5.2cm]{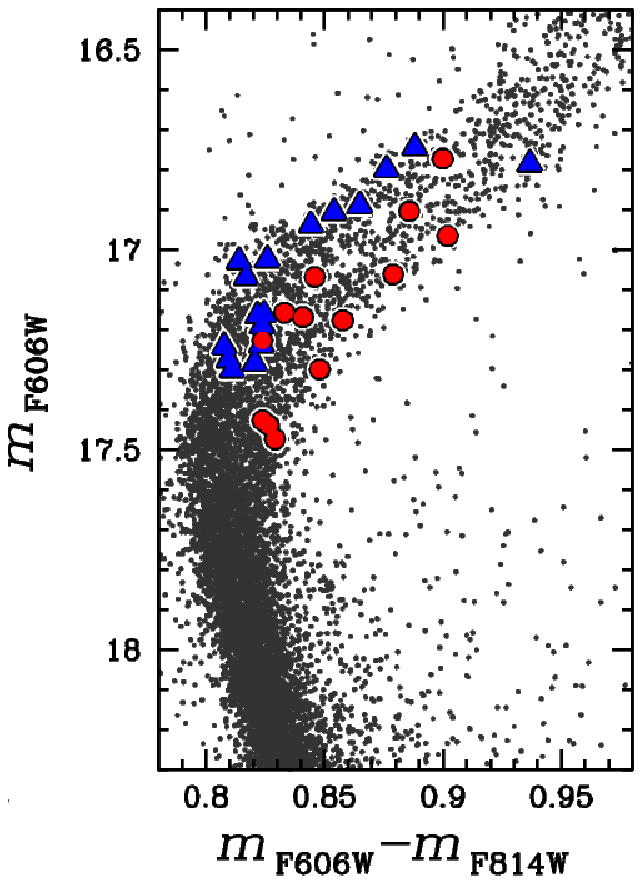}
\includegraphics[width=9cm,height=5cm]{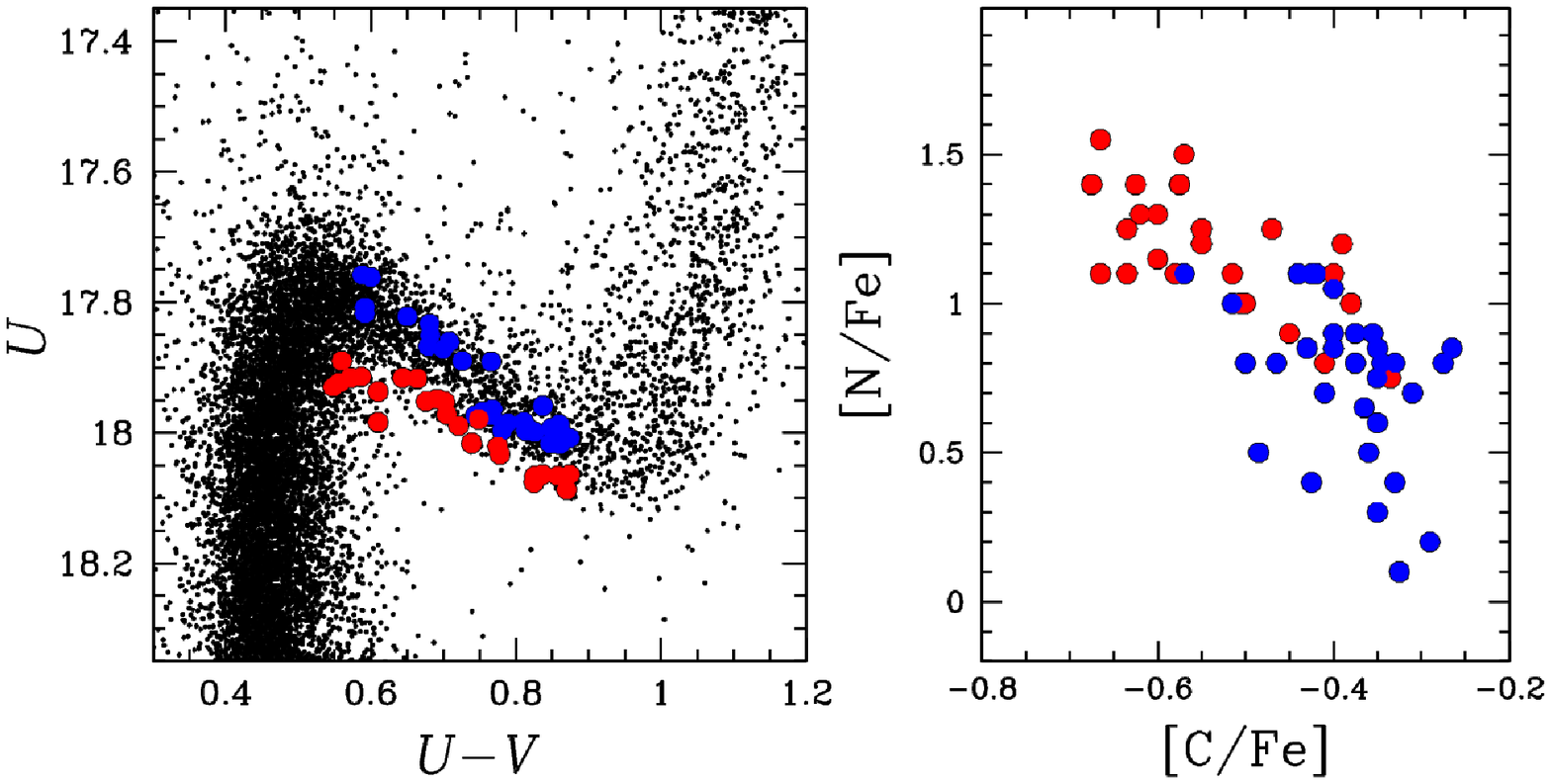}
\caption{\footnotesize
{\it Left panel}: M22 double SGB in the
$m_{\rm F606W}-m_{\rm F606W}$-$m_{\rm F814W}$ CMD (from \cite{mar12b}); {\it Right panels}:  $U$-$(U-V)$
CMD for 47~Tuc zoomed on the SGB. Stars in red and blue are our
spectroscopically analysed stars on the faint and bright SGB,
respectively.
The C and N abundances of these stars has been reported on the N-C
plane from \cite{mar13b}.
}
\label{cmd}
\end{figure*}
\end{center}

When appropriate colours, e.g $U-V$, are
used to construct the CMD, the two SGBs in M22
evolve in a double red-giant branch, associated with metallicity+CNO variations (\cite{mar11a,mar12b}).
A CMD showing the position of the M22 stars with
different $s$ content on the SGB has been shown in Fig.~\ref{cmd}
(left panel).
The fact that an {\it anomalously}
multi-modal SGB has been observed in the clusters with $s$ element variations,
suggets that the two phenomena may be
strictly connected, as higher $s$ abundances are correlated with C+N+O abundances.

A split SGB has been recently observed in the massive GC 47~Tuc
({\cite{and09}, \cite{mil12}, \cite{dic10}}, right panel in Fig~\ref{cmd}).
In this case there are no measurements of $n$-capture on SGB stars.
Preliminar results for chemical abundances in the light elements C and
N of SGB stars in 47~Tuc reveal that the fainter SGB hosts stars with
higher N content (Fig~\ref{cmd} from \cite{mar13b}).

\section{Conclusions}

We are discovering the presence of peculiarities in GCs, e.g. heavy element internal variations, complex (anti)correlations among light elements. 
Some GCs show variations in the chemical abundance of $n$-capture
elements due to enrichment from $s$-processes, in some cases, or
$r$-processes in other cases.
The variation in $s$-elements appears to be linked to peculiarities in
the CMDs, such as the SGB splits that have been associated with C+N+O
variations. The $r$-process enrichment in GCs seems to not be linked
to such peculiarities along the CMD.

Many issues are still contrived in depicting a picture for the formation of
multiple stellar populations, in particular for the {\it anomalous}
GCs, e.g.: 
{\it (i)}  the nature of the polluters for the $n$-capture enrichment; 
{\it (ii)} if an unique scenario can explain the heterogeneity of the multiple population zoo;
{\it (iii)} the possible extragalactic origin for {\it anomalous} GCs.

\begin{acknowledgements}
I am grateful to my collaborators for their help in the
publication of these results.
\end{acknowledgements}

\bibliographystyle{aa}

\end{document}